\documentclass[aps,
pra,
superscriptaddress,
twocolumn
]{revtex4-2}
\usepackage{braket}
\usepackage{quantikz}
\usepackage{graphicx}
\usepackage{float}
\usepackage{svg}
\usepackage{dcolumn}
\usepackage{bm, bbm}
\usepackage{lipsum}
\usepackage{amsthm}
\usepackage{siunitx}
\usepackage{physics}
\usepackage{amssymb}
\usepackage{placeins}
\usepackage{comment}
\numberwithin{equation}{section}
\usepackage[normalem]{ulem} 
\usepackage[colorlinks,linkcolor=blue,urlcolor=blue, citecolor=blue]{hyperref}
\usepackage{soul}

\def\captionof#1#2{{\def\@captype{#1}#2}}

\begin{document}

\preprint{APS/123-QED}

\title{Genuine Hybrid Number-Polarization Entanglement}

\author{Dorian Schiffer}
\email{Dorian.Schiffer@tuwien.ac.at}
\affiliation{Atominstitut, Technische Universit\"at Wien, Stadionallee 2, 1020 Vienna, Austria}
\affiliation{Institute for Quantum Optics and Quantum Information (IQOQI), Austrian Academy of Sciences,
Boltzmanngasse 3, 1090 Vienna, Austria}

\author{Marcus Huber}
\email{Marcus.Huber@tuwien.ac.at}
\affiliation{Atominstitut, Technische Universit\"at Wien, Stadionallee 2, 1020 Vienna, Austria}
\affiliation{Institute for Quantum Optics and Quantum Information (IQOQI), Austrian Academy of Sciences,
Boltzmanngasse 3, 1090 Vienna, Austria}

\author{Elizabeth Agudelo}
\email{Elizabeth.Agudelo@tuwien.ac.at}
\affiliation{Atominstitut, Technische Universit\"at Wien, Stadionallee 2, 1020 Vienna, Austria}
\affiliation{Institute for Quantum Optics and Quantum Information (IQOQI), Austrian Academy of Sciences,
Boltzmanngasse 3, 1090 Vienna, Austria}

\date{\today}

\begin{abstract}

Entanglement is a key resource for fundamental tests of physics and emerging quantum technologies. 
In quantum optics, two perspectives on entanglement coexist. 
In the continuous-variable framework, entanglement is understood as holding between optical modes. 
In contrast, discrete-variable quantum optics focuses on quantum correlations in degrees of freedom such as polarization that label fixed numbers of photons. In this paper, we show that entanglement can transcend this separation. 
Spontaneous parametric down-conversion inherently generates correlations in optical phase space, photon number, and labelling degrees of freedom simultaneously. 
In polarization, this structure is traditionally described by macroscopic Bell states. 
Existing witnesses, however, fail to detect the genuine hybrid entanglement of these states, which goes beyond the continuous-discrete-variable categorization. 
Here, we lay the groundwork for a general framework unifying continuous- and discrete-variable notions of entanglement. 
In particular, we derive an operational witness providing a sufficient criterion for genuine hybrid number-polarization entanglement and outline its experimental implementation. 
Finally, we discuss exemplary states which, together with our results on macroscopic Bell states, motivate a broader classification of genuine hybrid quantum correlations.
\end{abstract}

\maketitle

\section{Introduction}

Entanglement lies at the heart of quantum physics. 
Its counter-intuitive properties, challenging classical worldviews, led already Schrödinger to call it not only one, but \textit{the} characteristic trait of quantum mechanics \cite{Schroedinger1935}. 
Since then, entanglement has inspired investigations into the foundations of physics, both theoretical and experimental \cite{Einstein1935, Bell1964, CHSH1969, Freedman1972,Bell1980, Aspect1981, Aspect1982a, Aspect1982b, Bouwmeester1997, Weihs1998, Bertlmann2014, Giustina2015}, and the development of protocols that exploit it as a resource for communication and computation \cite{Bennett_Brassard1984, Ekert1991, Silberhorn2002, Josza2003, Groblacher2006, MS2000, Ding2007, Slussarenko2019}.\\

This long-standing interest has led to a wealth of methods for generating and detecting  entanglement across diverse quantum systems \cite{Peres1996, Horodecki1996, Duan2000, Simon2000, Giovannetti2003, Shchukin2005a, Shchukin2005b, Agarwal2005, Plenio2005,
GT09, HHHH09, Walborn2009, Sperling2009b,  Saboia2011, Tasca2013, Sperling2013, Hertz2016, Schneeloch2018, Friis2019, Floerchinger2021, Floerchinger2022, Gaerttner2023a, Gaerttner2023b}. 
And yet, even after quantum mechanics celebrated its first centennial, entanglement has not yielded all its secrets: Today, especially high-dimensional and multi-partite entanglement are areas of active research, promising advantages such as improved noise-resilience, computational speed-ups and many-party communication protocols \cite{Ecker2019, zhu2021, Erhard2020, Gao2023, Bulla2023, Pan2012}.\\

Paralleling the development of entanglement theory, light has remained a central platform for quantum physics. The generation of single photons \cite{Clauser1974}, squeezed light \cite{Stoler1970, Slusher1985, Lvovsky2016}, and photonic entanglement \cite{Wu1950, Kocher1967, Freedman1972, Aspect1981} marked major milestones in the maturation of modern quantum optics and photonics. 
Over time, this development branched into two communities. 
Continuous-variable (CV) quantum optics studies quantum effects in optical phase space and the underlying photon-number basis, whereas discrete-variable (DV) quantum optics focuses on degrees of freedom (DOFs) which are used to label certain fixed numbers of photons, such as polarization, spatio-temporal distribution, or frequency.
In the following, we collectively refer to these DOFs with the shorthand ``labelling DOFs''. 
It is well known that entanglement can exist between CV and DV systems \cite{vanLoock2011, Andersen2015}, where, for instance, one subsystem is described as a polarization qubit while the other is encoded in a CV state \cite{Andersen2010, Kwon2013, Morin2014, Jeong2014, Costanzo2015, Andersen2015, Gessner2016, Agudelo2017, Moradi2024}. 
Moreover, hybrid approaches combining CV and DV measurement techniques have been employed to demonstrate fundamental correlations \cite{Cavalcanti2011}.\\

These approaches, however, obscure the fact that the notion of entanglement differs drastically between CV and DV frameworks. 
In the CV setting, entanglement is understood as correlations between optical modes, manifesting in photon-number or quadrature correlations. 
In the DV framework, by contrast, entanglement is associated with correlations between labels of a fixed number of photons. 
This naturally raises the question whether entanglement can exist in a form that simultaneously shows CV and DV features.\\

Here, we answer this question affirmatively by demonstrating the hybrid nature of the entanglement generated in one of the most widespread technologies shared by both CV and DV quantum optics: \textit{Spontaneous parametric down-conversion} (SPDC) in non-linear crystals--- the workhorse behind many advances in both communities. In the CV context, SPCD is the source of single- and two-mode squeezed states (TMSV) which are the quintessential CV resources in quantum optics and CV quantum information processing \cite{Braunstein2005, Weedbrook2012, Lvovsky2016, Usenko2025}.
In the DV framework, for example, sources of polarization entanglement typically require the interference of at least two SPDC events \cite{Anwar2021}. 
Within this community, the source is commonly described as producing entangled photon pairs.
However, this picture is merely an approximation. 
The actual output state in Fock space \cite{Fock1932} is a coherent superposition of infinitely many polarization-entangled multi-pair contributions, with amplitudes determined by the pump strength. 
Because these states describe polarization-entangled light beams rather than two entangled photonic qubits, resulting in macroscopically distinguishable quantum states \cite{Chekhova2015}, they are commonly referred to as \textit{macroscopic Bell states} (MBS).\\

Owing to the MBS structure, strongly pumped SPDC-based polarization-entanglement sources exhibit correlations not only in polarization but also in Fock space and optical phase space. 
Thus, even these commonplace states transcend the conventional dichotomy between CV and DV entanglement, pointing towards a previously overlooked class of quantum correlations. 
Although Bell-type inequalities \cite{Huntington2001, Howell2002, Thearle2018, Oliveira2024} and entanglement witnesses \cite{Simon2003, Eisenberg2004, Stobinska2012, Kanseri2013} have been employed to detect entanglement in MBS, these approaches rely exclusively on polarization correlations and therefore fail to capture the \textit{genuine hybrid number-polarization entanglement} inherent to these sources. 
Correspondingly, early proposals employing MBS in quantum information protocols such as quantum key distribution likewise exploited only their polarization entanglement \cite{Durkin2002, Luo2005}.\\

In this paper, we close the conceptual gap between CV and DV entanglement and derive a meaningful witness that yields a rigorous, sufficient criterion for genuine hybrid number-polarization entanglement.
We first review the MBS formalism and existing entanglement witnesses. 
We then use MBS as a reference to derive an operationally meaningful witness for genuine hybrid number-polarization entanglement and outline its experimental implementation. Importantly, MBS are not the only states that evade the standard CV-DV classification. 
We therefore go beyond this specific example and discuss a broader class of states whose entanglement lies outside the conventional CV and DV frameworks altogether. Notice, however, that our considerations of CV and DV entanglement are fundamentally different from the rigorous distinctions made between particle and field entanglement in the contexts of first and second quantization \cite{Sperling2023}.\\

Against the backdrop of the ubiquity of SPDC sources, together with the growing interest in entanglement in high-dimensional, multipartite, and hybrid quantum systems connecting CV and DV regimes, our demonstration of hybrid correlations in MBS opens the door to their future integration into protocols that bridge CV and DV paradigms. 
More broadly, our results suggest that hybrid entanglement is a natural unifying framework between discrete- and continuous-variable quantum optics.

\section{Macroscopic Bell States}

Experimentally, MBS are prepared by first generating two TMSVs from a strong coherent pump undergoing unseeded SPDC in a non-linear crystal and subsequently overlapping both states, for instance, in a Sagnac configuration \cite{Shi2004, Kim2006}. 
Formally, we represent this situation by considering a four-mode Fock space, where two parties, Alice and Bob, each hold two bosonic modes with orthogonal polarizations, which we shall call $H$ and $V$. 
In this framework, the unitaries generating macroscopic Bell states are given by
\begin{align}
     \mathcal{S}_\Phi^\pm &= \exp \left[ \xi^* (a_H b_H \pm a_V b_V) - \text{H. c.}\right] \text{ and }\label{eq:S_phi}\\
      \mathcal{S}_\Psi ^\pm &= \exp \left[ \xi^* (a_H b_V \pm a_V b_H) - \text{H. c.}\right]. \label{eq:S_psi}
\end{align}
Here, $a_k, b_k$ with $k \in \{H,V\}$ are the annihilation operators associated with the modes at Alice and Bob, respectively. 
The squeezing parameter $\xi = re^{i\theta}$ is related to the pump, the length of the crystal, and the strength of its second-order nonlinearity. 
The MBS are defined as
\begin{align}
    \ket{\Phi^\pm} &\equiv \mathcal{S}_\Phi^\pm \ket{0000},\\
    \ket{\Psi^\pm} &\equiv  \mathcal{S}_\Psi ^\pm \ket{0000}.
\end{align}
We denote the photon numbers per mode as $\ket{N^A_H, N^A_V, N^B_H, N^B_V}$, where the first two entries relate to Alice and the last two to Bob. 
Using the number-basis representation of TMSV states, the macroscopic singlet reads
\begin{equation}
   \ket{\Psi^-}  = \sum_{n,m = 0}^{\infty} (-1)^m \frac{\tanh(r)^{n+m}}{\cosh^2(r)} \ket{nmmn},
    \label{eq:Psi_Mac}
\end{equation}
for $\theta = 0$. 
Closer inspection shows that MBS can be viewed as two TMSV states whose modes are appropriately re-routed to generate polarization entanglement. 
Rewriting the state \cite{Simon2003, Stobinska2012} makes this explicit:
\begin{align}
   \ket{\Psi^-}   &= \sum_{n =0}^\infty f_n\ket{nn}_{A_H, B_V}  \otimes \sum_{m =0}^\infty f_m\ket{mm}_{A_V, B_H}\\  
    &= \sum_{N=0}^\infty \sum_{\substack{n,m = 0 \\n+m = N}}^\infty f_{n,m}\ket{nmmn}\\
    &=  \sum_{N=0}^\infty \sum_{n = 0}^N f_{n, N-n}\ket{n,N-n,N-n,n},
\end{align}
where $f_{n,m}$ denotes the expansion coefficients from Eq.~\eqref{eq:Psi_Mac} and $N$ counts the local photon number across polarization modes. 
Plugging in the explicit form of $f_{nm}$ and normalizing, we obtain
\begin{equation}
     \ket{\Psi^-} = \frac{1}{\cosh^2(r)} \sum_{N=0}^\infty \tanh(r)^{N}\sqrt{N+1}\ket{\psi_N^-},
     \label{eq:mac_singlet}
\end{equation}
with
\begin{equation}
    \ket{\psi_N^-} = \frac{1}{\sqrt{N+1}} \sum_{n = 0}^N (-1)^{(N-n)}\ket{n,N-n,N-n,n}.
    \label{eq:per_sector_states}
\end{equation}\\

The states $\ket{\psi_N^-}$ are equivalent to singlets of two effective spin-$N/2$ systems \cite{Simon2003}. 
Hence, MBS has been characterized as the singlet state of a large but highly uncertain spin \cite{Stobinska2012}. 
This can be motivated by introducing \textit{Stokes operators},
\begin{align}
    S_x &= \frac{1}{2} (a^\dagger_D a^{~} _D - a^\dagger_A a^{~}_A ),\\
    S_y &= \frac{1}{2} (a^\dagger_L a^{~}_L - a^\dagger_R a^{~}_R),\\
    S_z &= \frac{1}{2} (a^\dagger_H a^{~}_H  - a^\dagger_V a^{{~}}_V ).
\end{align}
Here, the indices $D,A$ and $R,L$ relate to diagonally and anti-diagonally and to right and left circular polarized light. 
States $\ket{\psi_N^-}$ are the singlet states for the total angular momentum operator,
\begin{equation}
    \textbf{S}= \textbf{S}_A + \textbf{S}_B,
\end{equation}
where $\textbf{S}_A = (S_x^A, S_y^A, S_z^A)^T$ and similarly for Bob, for the total spin of $N/2$. This entails polarization entanglement between Alice and Bob. 
In particular, $\bra{\Psi^-}\textbf{S}^2\ket{\Psi^-} = 0$ implies that MBS exhibit quantum correlations in the Stokes operators together with the reduction of polarization noise \cite{Korolkova2002, Iskhakov2011a, Iskhakov2011b, Kanseri2013, Prasannan2022}.\\

These polarization correlations have been used to witness entanglement in MBS. For example,in Ref.~\cite{Simon2003}  a lower bound on $\langle\textbf{S}^2 \rangle$ for separable states was derived, which is violated by macroscopic singlets, an approach that was extended in Ref.~\cite{Stobinska2012} to the other MBS using the variances of Stokes operators. 
Alternatively, researchers had followed the results presented in Ref.~\cite{Huntington2001} and directly tested Stokes-operator correlations via a Clauser-Horne-Shimony-Holt (CHSH) inequality \cite{CHSH1969}.\\

However, these techniques target only polarization entanglement within each fixed-$N$ sector. 
They fail to detect the number-basis entanglement of MBS. 
In fact, the coherent superposition of $\ket{\psi_N^-}$ that defines MBS is not required to trigger the above methods. 
For suitable weights $w_N$, even the classical mixture
\begin{equation}
    \rho = \sum_{N = 0}^\infty w_N\ket{\psi_N^-}\bra{\psi_N^-}
    \label{eq:psiN_mix}
\end{equation}
will reproduce the same violations. 
Fundamentally, this limitation arises from the choice of observables these witnesses employ: Stokes operators and their variances commute with the total number operator. 
Thus, they access only the block-diagonal part of the Fock-space density matrix and are insensitive to the coherences across photon-number sectors. 
Hence, these existing strategies do not capture the genuine hybrid number-polarization entanglement we unveil in this work.

\section{State Space and Witness}
To introduce genuine hybrid number-polarization entanglement and derive a corresponding witness, we first describe the state space under consideration. Let us define three convex sets of states,
\begin{align}
    \mathrm{Sep} &\coloneq \{ \rho~|~\rho \text{ n-separable } \land \text{ p-separable}\},\\
    \mathrm{N} &\coloneq \{ \rho~|~\rho \text{ n-separable}\},\\
    \mathrm{P} &\coloneq \{ \rho~|~\rho \text{ p-separable}\}.
\end{align}
While all these sets are defined by types of separability, they differ formally: Set Sep contains states that are fully n(umber)- and p(olarization)-separable. 
Only these states can straightforwardly be understood as convex mixtures of product states. 
In contrast, states of set N may not be represented as a product state. They are instead characterized by the absence of quantum correlations in photon number. 
Note that this does not exclude DV entanglement, for example, in polarization, which is constrained to a fixed-$N$ Fock sector. 
This implies that the latter can always be viewed as quantum correlations between labels of a fixed number of photons or convex mixtures of such. 
Specifically, the state \eqref{eq:psiN_mix} is located in set N, which shows no number entanglement in this sense.\\

In contrast, set P consists of states that are separable within each Fock sector. 
Here, post-selection on a fixed photon number will yield product states. 
Again, this definition allows for number or CV entanglement. 
Consequently, the paradigmatic example for set P is the TMSV state, which describes the output state of an individual non-degenerate SPDC event. 
For a TMSV in modes labelled by DOFs $a$ and $b$, the state of any $N$-photon sector will be of the form $\ket{a}^{\otimes N/2}\otimes \ket{b}^{\otimes N/2}$, such that no quantum correlations in the labelling DOF occur.\\

The convex hull of N and P, $\text{conv}(\mathrm{N}\cup\mathrm{P})$, contains on the one hand the fully separable states, since $\mathrm{Sep} = \mathrm{N} \cap \mathrm{P}$, and on the other hand, a class of states that might exhibit number entanglement and polarization entanglement. 
Despite this, these states are merely mixtures of states from N and P. 
Beyond $\text{conv}(\mathrm{N}\cup\mathrm{P})$, we find set Hyb, which is composed of states that are \textit{genuinely} hybrid number-polarization entangled.\\

\begin{figure}
\vspace{0.5em}
\begin{center}
\includegraphics[width=0.7\columnwidth]{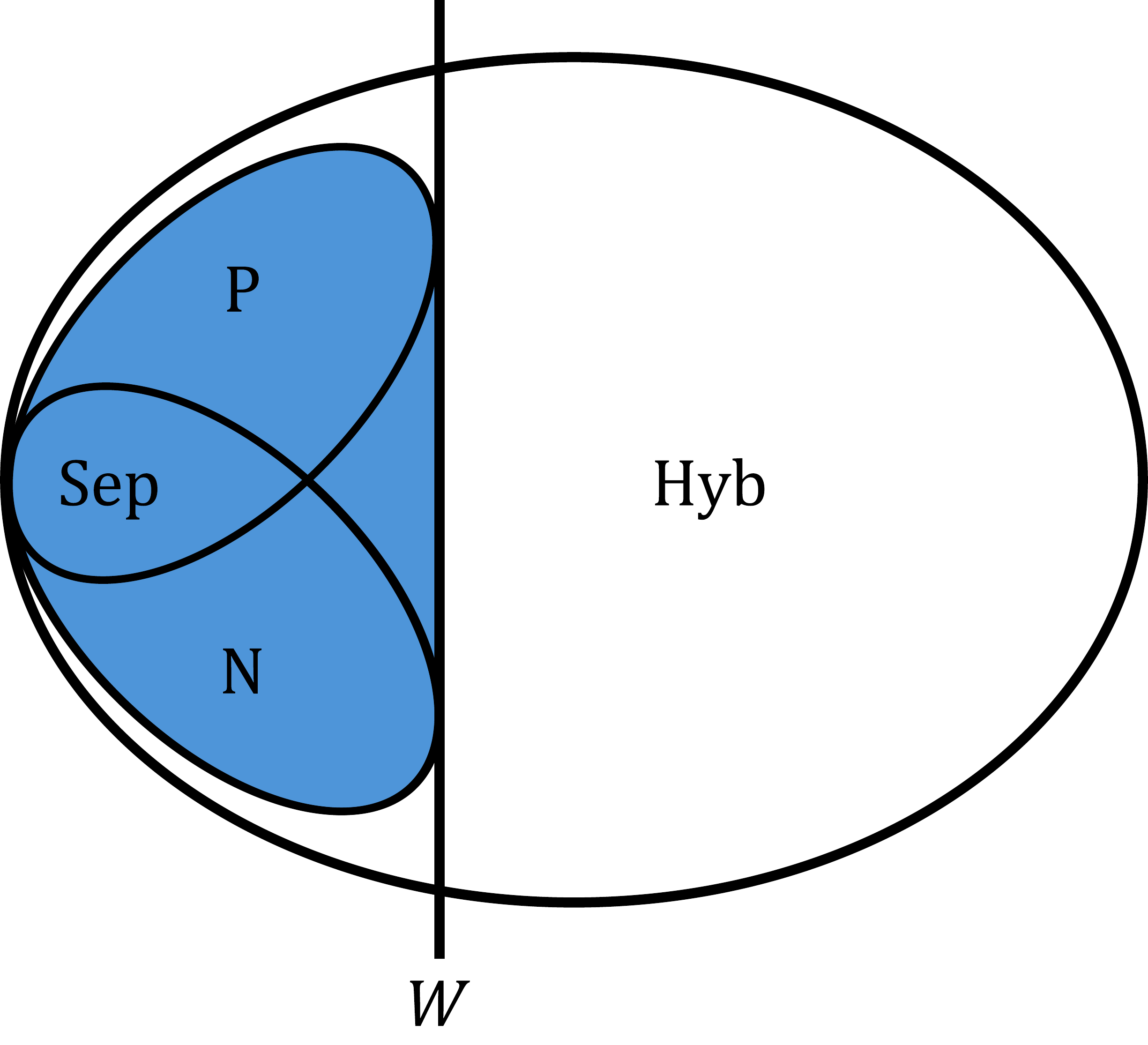}
\end{center}
\caption{State space and proposed witness. 
The space of fully separable states, $\mathrm{Sep}$, is the intersection of the polarization-separable states, $\mathrm{P}$, and of the number-separable states, $\mathrm{N}$. 
Beyond these partially separable states lies the set of genuinely hybrid number-polarization-entangled states, $\mathrm{Hyb}$. 
Witness $W$ constitutes a supporting hyperplane to the convex hull of N and P drawn in blue.} 
\label{fig:state_space}
\end{figure}

Existing entanglement witnesses \cite{Huntington2001,Simon2003,Eisenberg2004,Stobinska2012}, however, do not distinguish between states $\rho \notin \mathrm{P}$ and states in Hyb, such that polarization entanglement is sufficient to trigger a violation. 
This implies that to detect genuine hybrid number-polarization entanglement for a state $\rho$, we need to show $\rho \notin \text{conv}(\mathrm{N}\cup\mathrm{P})$. 
Figure \ref{fig:state_space} illustrates the state space defined by the sets introduced above, as well as the hyperplane defined by the desired witness operator $W$.\\

We argue that MBS are not only contained in Hyb but in fact constitute prototypical representatives of this type of states. 
This directly follows from their structure: MBS are constructed from TMSVs, which exhibit optimal photon-number correlations while simultaneously being maximally polarization-entangled within each fixed Fock sector, i.e.,
\begin{equation}                                            \text{Tr}_B \left( \ket{\psi_N^-}\bra{\psi_N^-} \right) = \frac{1}{N+1} \ \mathbbm{1}_N,
\end{equation}
where $\mathbbm{1}_N$ is the identity operator on the $N+1$-dimensional Fock sector. 
Taken together, these features show that MBS simultaneously saturate correlations in photon number and polarization, placing them squarely within Hyb. 
Therefore, states in the vicinity of MBS share this hybrid structure. 
This motivates using the \textit{fidelity} to MBS as a witness for genuine hybrid number–polarization entanglement.\\

Drawing on the canonical form of fidelity-based witnesses \cite{Sperling2009, Friis2019}, we formulate the witness operator
\begin{equation}
    W = \mathcal{F}\mathbbm{1} - \ket{\Psi^-}\bra{\Psi^-},
    \label{eq:hybrid_witness}
\end{equation}
where $\mathcal{F} \in \mathbbm{R}$ and $\mathbbm{1}$ denotes the identity operator on the four-mode Fock space. 
Given that MBS are equivalent under local operations and classical communication (LOCC), we again choose without loss of generality the macroscopic singlet $\ket{\Psi^-}$ to represent all MBS. 
We then require that $\langle W \rangle = \Tr(W\rho) \geq 0$ holds for all states within the convex hull of N and P, such that
\begin{equation}
    \langle W \rangle < 0
    \label{eq:criterion}
\end{equation}
signifies genuine hybrid number–polarization entanglement. 
To this end, we define
\begin{equation}
    \mathcal{F}_W \coloneq \sup_{\rho \in \text{conv}(\mathrm{N}\cup\mathrm{P})}\Tr\left(\rho \ket{\Psi^-}\bra{\Psi^-}\right),
\end{equation}
and the suprema
\begin{align}
     \mathcal{F}^\mathrm{N}_\text{sup} &\coloneq \sup_{\rho \in \mathrm{N}}\Tr\left(\rho \ket{\Psi^-}\bra{\Psi^-}\right),\\
     \mathcal{F}^\mathrm{P}_\text{sup} &\coloneq \sup_{\rho \in \mathrm{P}}\Tr\left(\rho \ket{\Psi^-}\bra{\Psi^-}\right).
\end{align}
Recall that we can decompose any $\rho \in \text{conv}(\mathrm{N}\cup\mathrm{P})$ as
\begin{equation}
    \rho = \lambda \rho_\mathrm{N} + (1- \lambda) \rho_\mathrm{P},
\end{equation}
where $0 \leq \lambda \leq 1$. 
From the linearity of the overlap and the convexity of the above decomposition, we conclude that for any $\rho$ in the convex hull of N and P, we have
\begin{align}
    \Tr\left(\rho \ket{\Psi^-}\bra{\Psi^-}\right) &= \lambda  \mathcal{F}^\mathrm{N} + (1- \lambda) \mathcal{F}^\mathrm{P}\\
    &\leq \max( \mathcal{F}^\mathrm{N},\mathcal{F}^\mathrm{P})\\
    &\leq \max( \mathcal{F}^\mathrm{N}_\text{sup},\mathcal{F}^\mathrm{P}_\text{sup}). \label{eq:bound_derivation}
\end{align}
Here, $\mathcal{F}^\mathrm{N} = \Tr(\rho_\mathrm{N} \ket{\Psi^-}\bra{\Psi^-})$, and similarly for set P, denote the individual overlaps of states contained in N and P with MBS respectively.\\

Since the bound \eqref{eq:bound_derivation} holds for every element of the convex hull, we conclude that
\begin{equation}
     \mathcal{F}_W \leq \max \left( \mathcal{F}^\mathrm{N}_\text{sup}, \mathcal{F}^\mathrm{P}_\text{sup} \right).
     \label{eq:F_sep}
\end{equation}
It follows that we may construct the sought-after witness by choosing $\mathcal{F} = \max( \mathcal{F}^\mathrm{N}_\text{sup}, \mathcal{F}^\mathrm{P}_\text{sup})$ back in \eqref{eq:hybrid_witness}. 
Notice that, in the presence of noise, losses, or other experimental imperfections, even states within Hyb might fail to violate this witness. 
Consequently, the criterion \eqref{eq:criterion} is only \textit{sufficient}.\\

To render our witness explicit, we derive upper bounds to the suprema appearing in \eqref{eq:F_sep}. Due to the linearity of the objective function and the convexity of N and P, these will be attained by pure states. 
For bipartite pure states, the presence of non-trivial correlations is equivalent to entanglement.\\

For the number-separable states, we hence restrict ourselves to pure states without correlations in the number basis. 
These are, on the one hand, product states, which are, however, in Sep, and, on the other hand, states with arbitrary structure within fixed-$N$ subspaces, but no coherences across Fock sectors, i.e., block-diagonal states. 
Such states might still be maximally polarization entangled for any fixed photon number and thereby achieve unit overlap with MBS for a certain Fock sector. 
Hence,
\begin{equation}
    \mathcal{F}^\mathrm{N}_\text{sup} \leq \max_{N \in \mathbbm{N}} p_N, 
    \label{eq:F_N}
\end{equation}
where
\begin{equation}
    p_N = \frac{\tanh(r)^{2N}}{\cosh^4(r)}(N+1)
\end{equation}
are the probabilities computed from \eqref{eq:mac_singlet} for the $N$-photon Fock sector to be occupied. For polarization-separable states, separability must hold within each Fock sector. 
As we show in the Appendix, this entails
\begin{equation}
    \mathcal{F}^\mathrm{P}_\text{sup} \leq \sum_{N = 0}^\infty\frac{p_N}{N+1}.
    \label{eq:F_P}
\end{equation}
Together, \eqref{eq:F_N} and \eqref{eq:F_P} formally conclude the construction of the witness. 
However, the full domains of maximization and summation are not experimentally accessible. 
Below, we will discuss how to implement our method in a lab.

\section{Discussion}
In this section, we first outline the experimental implementation of the witness. 
We then turn to the broader implications of our work by analysing exemplary states obtained from an alternative decomposition of MBS and number state interference at a beam splitter.\\

In standard fidelity-based entanglement witnesses \cite{Friis2019}, the reference or target state should be chosen as close as possible to the experimentally produced state to maximize the witness performance. An imperfect choice of target state does not invalidate any detected entanglement, but reduces the likelihood of observing a violation. For MBS, the reference states are parametrized solely by the squeezing parameter $r$. Consequently, the separability bounds derived above depend explicitly on the squeezing. To assess whether a laboratory state exhibits genuine hybrid number-polarization entanglement, one therefore first estimates its squeezing, for instance via the second moments of the involved quadratures. This experimentally determined squeezing is then used to select the corresponding MBS reference state and evaluate the separability bounds.\\

Once the squeezing of the state in question has been obtained, its overlap with MBS must be computed. 
This requires tomography of the Fock-state density matrix, for example by means of \textit{pattern functions} \cite{Leonhardt1995a, Leonhardt1995b, Lvovsky2009}. 
In practice, however, the density matrix can be reconstructed only up to finite photon numbers. 
Recall that local filtering of the photon number is LOCC and cannot create entanglement. We therefore may evaluate the bounds \eqref{eq:F_N} and \eqref{eq:F_P} within the experimentally accessible subspace, using the renormalized probabilities $\Tilde{p}_N$. 
The resulting witness yields a sufficient criterion for genuine hybrid number-polarization entanglement for the state conditioned on this subspace.\\

\begin{figure}
\begin{center}
\includegraphics[width=\columnwidth]{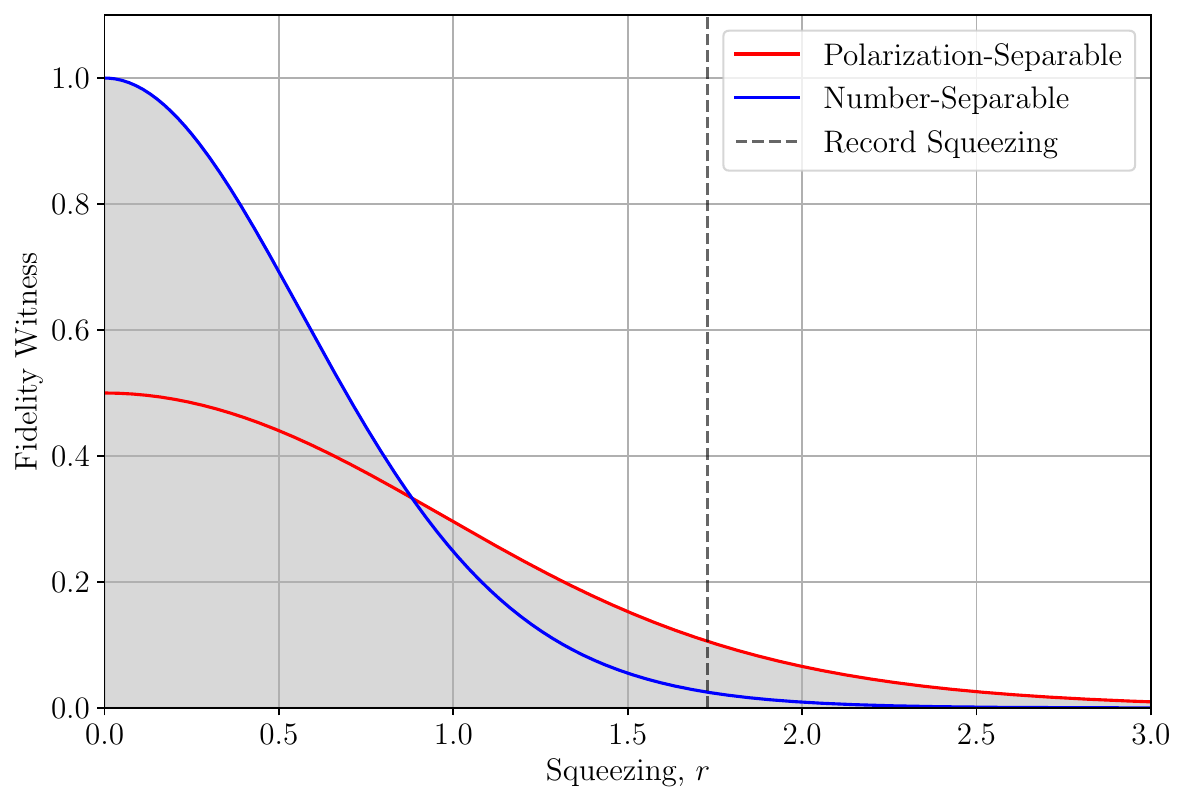}
\end{center}
\caption{Scaling of the hybrid entanglement witness versus squeezing in the non-vacuum sector of Fock space. 
The fidelity to MBS needs to clear the grayed-out area in order to certify the presence of genuine hybrid number-polarization entanglement. 
The dashed line indicates the current world record for measured squeezing, achieved by the authors of Ref. \cite{Vahlbruch2016}, demonstrating the experimental feasibility of our technique.}
\label{fig:witness_vs_squeezing}
\end{figure}

For illustration, Figure \ref{fig:witness_vs_squeezing} shows the scaling of the witness evaluated in the non-vacuum subspace of Fock space, i.e., $N \geq 1$, as a function of the squeezing parameter. 
Restricting oneself to the subspace in which entanglement can actually occur seems like a natural choice for experimental implementations. 
Notice that the boundary drawn for number separability is the overlap of the two-photon singlet. 
At high squeezing, approximately above $r = 1.15$, however, the overlap with the four-photon singlet will be slightly larger and thereby set the number separability bound. 
Operationally, this transition is irrelevant, as in this regime, the dominant bound is polarization separability.\\

We briefly comment on the experimental feasibility of our approach. 
For once, the level of squeezing required to potentially violate our witness lies well below the current world record for squeezed light \cite{Vahlbruch2016}, indicated as a dashed line in Figure \ref{fig:witness_vs_squeezing}. 
Moreover, Ref.~\cite{Thearle2018} reports MBS generation with a squeezing of $r \approx 0.5$, for which the fidelity required to demonstrate genuine hybrid entanglement, around $0.65$, is moderate. 
These observations suggest that our method is within reach of state-of-the-art experiments.\\

Finally, despite standing at the centre of this paper, polarization is only one possible internal DOF for realizing MBS. 
Considering SPDC Hamiltonians that involve DOFs like time, energy, or spectral modes \cite{Christ2011, Christ2012}, we anticipate a similar structure, albeit with high-dimensional entanglement already in the single photon-pair Fock sector. 
In these situations, the coherent overlap of TMSVs is achieved not by manually interfering SPDC events, as in Sagnac-type schemes and related strategies, but by transferring, for example, temporal coherence from the pump to the generated state. 
Expanding the MBS formalism to such high-dimensional labelling DOFs is left as an interesting avenue for future work.\\

We now turn to the implications of hybrid number-polarization entanglement for the landscape of entanglement in quantum optics. 
As we have seen, the standard distinctions between entanglement shared by photons and entanglement shared by optical modes are both simultaneously applicable to MBS. 
This suggests that entanglement itself may be more fundamental, beyond the conventional CV-DV divide. This point can be illustrated further through exemplary states. 
Consider again the representation of MBS given in Eq.~\eqref{eq:Psi_Mac}. 
Previously, we grouped terms according to fixed total photon number, yielding the $N$-sector decomposition and thereby revealing the singlet structure. 
However, one may equally choose a decomposition that emphasizes superpositions across both Fock sectors and polarization modes. 
Restricting ourselves to the single- and double-pair contributions, we define the \textit{cross-layer state}
\begin{equation}
    \ket{\mathcal{C}} = \alpha \ket{2002} + \beta \ket{0110},
\end{equation}
where $\alpha, \beta \in \mathbb{C}$ are suitably normalized amplitudes. 
This state is clearly entangled. 
However, the correlations connect a two-photon component with a one-photon component, preventing an interpretation in terms of standard DV entanglement, which presupposes an equal particle number. 
At the same time, the state does not exhibit the characteristic photon-number correlations associated with TMSV states and therefore also resists a straightforward classification as CV entanglement. Although such states appear only coherently embedded within MBS and cannot easily be isolated experimentally, their puzzling structure nevertheless underscores the necessity of a more general notion of entanglement. To conclude with a more experimentally accessible example, consider interfering a $\ket{2_H 1_V}$-state in a single spatial mode on a beam splitter. 
Routing one output port to Alice and the other to Bob, while keeping the mode labelling as introduced above, one may post-select the \textit{beam-splitter state} given by
\begin{equation}
    \ket{\mathcal{B}} = \alpha \ket{2001} + \beta \ket{0120}.
\end{equation}
While the symmetry of this state differs from cross-layer states, it again gives rise to correlations that do not naturally fit within the standard CV or DV paradigms.\\

Together with the main result of this paper, these examples indicate that we may conceive states that may simultaneously exhibit both CV and DV entanglement, or, alternatively, evade both classifications when only common CV or DV correlations are considered. 
Nevertheless, such states remain genuinely entangled within a more complete hybrid framework. 
The development of such a framework is equally left for future work.

\section{Conclusions}
In this paper, we have demonstrated entanglement beyond the conventional dichotomy between discrete-variables and continuous-variables in quantum optics. 
An important family of states exhibiting such correlations are macroscopic Bell states generated by strongly pumped spontaneous-parametric-down-conversion sources. 
Previous attempts to capture their entanglement failed to detect their hybrid nature.\\

By using macroscopic Bell states as a reference that saturates these hybrid correlations, we derived a fidelity-based witness providing a sufficient criterion for genuine hybrid number-polarization entanglement. 
We further outlined its experimental implementation and the evaluation of the relevant bounds. 
Recent experiments have shown the preparation of macroscopic Bell states at the required squeezing levels, indicating that the violation of our witness is experimentally achievable. 
Our technique is therefore a directly applicable method to certify genuine hybrid number-polarization entanglement, constituting the first step towards potential exploitation of this novel quantum resource. Finally, through additional examples, we discussed the broader consequences of genuine hybrid entanglement for the distinction between continuous-variable and discrete-variable entanglement.\\

Much like in multi-partite systems, where entanglement across every bipartition does not imply genuine multipartite entanglement, we have introduced genuine hybrid number-polarization entanglement. States with this property cannot be written as classical mixtures of purely CV- or DV-entangled states, revealing a form of entanglement beyond either description. These insights motivate the development of a general framework for entanglement that transcends these categories.

\section*{Acknowledgements}
D.S.~is grateful to Florian Kanitschar for many fruitful discussions. 
M.H.~acknowledges funding from the European Research Council (Consolidator grant `Cocoquest' 101043705). E.A.~acknowledges funding from the Austrian Science Fund (FWF) through the Elise Richter project HyDRA 10.55776/V1037.
Views and opinions expressed are, however, those of the authors only and do not necessarily reflect those of the funding agencies. 
Neither can the funding agencies be held responsible for them.

\bibliography{sources.bib}

\newpage
\onecolumngrid

\appendix
\section{Proof of the Polarization-Separability Bound}

Here, we derive the upper bound to the overlap any state $\rho \in \mathrm{P}$ can have with MBS. In contrast to the number-separability bound, we now cannot bring the block-diagonal structure to bear, because $\rho_\mathrm{P}$ may have arbitrary number entanglement. We begin with deriving in general
\begin{align}
    \Tr\left(\rho \ \ket{\Psi^-} \bra{\Psi^-}\right) &= \sum_{N=0}^\infty |f_N|^2\bra{\psi_N^-} \rho \ket{\psi_N^-} + \sum_{\substack{N, M=0 \\ N \neq M}}^\infty f_N  f^*_M \bra{\psi_M^-} \rho \ket{\psi_N^-}\\
     &= \sum_{N=0}^\infty |f_N|^2\bra{\psi_N^-} \rho \ket{\psi_N^-} + 2\sum_{\substack{N, M=0 \\ N  >M}}^\infty \Re(f_N  f^*_M \bra{\psi_M^-} \rho \ket{\psi_N^-})\\
     &\leq \sum_{N=0}^\infty |f_N|^2\bra{\psi_N^-} \rho \ket{\psi_N^-} + 2\sum_{\substack{N, M=0 \\ N  >M}}^\infty \sqrt{\Re(f_N  f^*_N \bra{\psi_N^-} \rho \ket{\psi_N^-})\Re(f_M  f^*_M \bra{\psi_M^-} \rho \ket{\psi_M^-})}\\
     &=\left( \sum_{N=0}^\infty \sqrt{ |f_N|^2\bra{\psi_N^-} \rho \ket{\psi_N^-}} \right)^2,
\end{align}
where $f_N$ denote the expansion coefficients as given in \eqref{eq:mac_singlet}. We optimize over the linear target function over the convex set P and hence choose $\rho_\mathrm{P} = \ket{\Phi_\mathrm{P}} \bra{\Phi_\mathrm{P}}$, anticipating that the pure state
\begin{equation}
    \ket{\Phi_\mathrm{P}} = \sum_{K =0}^\infty c_K \ket{\phi_K},
\end{equation}
which describes a coherent superposition with normalized amplitudes $c_K \in \mathbb{C}$ of the polarization-separable states $\ket{\phi_K}$ located in the $K$-local-photons sector, will yield the desired bound. Formally, the polarization separability requires
\begin{equation}
    \Tr \left( \ket{\phi_K}\bra{\phi_K} \ \ket{\psi^-_K}\bra{\psi^-_K}\right) \leq \frac{1}{K+1}.
    \label{eq:partsep}
\end{equation}
Notice that depending on the $c_k$, state $\ket{\Phi_\mathrm{P}}$ is either fully separable or number entangled. We plug in and find
\begin{equation}
    \Tr\left(\rho_\mathrm{P} \ \ket{\Psi^-} \bra{\Psi^-}\right) \leq \left( \sum_{N=0}^\infty \sqrt{ |f_N|^2\bra{\psi_N^-} \rho_\mathrm{P} \ket{\psi_N^-}} \right)^2 = \left( \sum_{N=0}^\infty \sqrt{  \sum_{K, L=0}^\infty c_Kc_L^*|f_N|^2\bra{\psi_N^-}  \phi_K \rangle\langle\phi_L \ket{\psi_N^-}} \right)^2.
\end{equation}
Since $\bra{\psi_N^-}  \phi_K \rangle = 0$ if $N\neq K$, this sum collapses to
\begin{equation}
    \Tr\left(\rho_\mathrm{P} \ \ket{\Psi^-} \bra{\Psi^-}\right) \leq  \left( \sum_{N=0}^\infty \sqrt{|c_N|^2 \ |f_N|^2 \  |\bra{\psi_N^-}  \phi_N \rangle|^2} \right)^2 \leq  \left( \sum_{N=0}^\infty \sqrt{|c_N|^2 \ |f_N|^2 \ \frac{1}{N+1}} \right)^2.
\end{equation}
We can now upper bound this further by using the Cauchy-Schwarz inequality via
\begin{equation}
   \Tr\left(\rho_\mathrm{P} \ \ket{\Psi^-} \bra{\Psi^-}\right) \leq\left( \sum_{N=0}^\infty \sqrt{  |c_N|^2 \ |f_N|^2 \frac{1}{N+1}} \right)^2 \leq \left(\sum_{N = 0}^\infty\frac{|f_N|^2}{N+1}\right) \left(\sum_{N = 0}^\infty|c_N|^2\right) =  \sum_{N = 0}^\infty\frac{p_N}{N+1}
\end{equation}
Evaluating the bound on the full Fock space yields $1/\cosh^2(r)$, which is attained by the state $\rho^\text{TMSV}_{A_H, B_V}\otimes \ket{00}\bra{00}_{A_V, B_H}$. However, here the vacuum artificially enhances the overlap of any state in P such that the bound evaluated over N may never beat it. This seems undesirable, which is why in Fig.~\ref{fig:witness_vs_squeezing} we plot the bound evaluated in the non-vacuum subsector, where we evaluate the witness using renormalized probabilities, $\Tilde{p}_N = p_N/(1-p_0)$, and find that
\begin{equation}
     \sum_{N = 1}^\infty\frac{\Tilde{p}_N}{N+1} = \frac{2}{\cosh(2r)+3}.
\end{equation}
Following this procedure, the bounds may be evaluated in any accessible subspace. Inspection of a few low-photon-number examples reveals that the bounds scale comparably to the ones we plot in the main body of this paper, ensuring that violation is still feasible with experimentally accessible squeezing.

\end{document}